\documentclass[lettersize,journal]{IEEEtran}
\usepackage{amsmath,amsfonts,amssymb}
\usepackage{algorithmic}
\usepackage{array}
\usepackage{graphicx} 
\usepackage[caption=false,font=footnotesize]{subfig}

\usepackage{textcomp}
\usepackage{stfloats}
\usepackage{url}
\usepackage{verbatim}
\usepackage{cite}
\usepackage{xcolor}
\usepackage{comment}
\usepackage{mathrsfs}
\usepackage{upgreek}
\usepackage{amsthm}
\usepackage{multirow}

\newtheorem{definition}{Definition}

\newtheorem{lemma}{Lemma}

\def\BibTeX{{\rm B\kern-.05em{\sc i\kern-.025em b}\kern-.08em
    T\kern-.1667em\lower.7ex\hbox{E}\kern-.125emX}}
\usepackage{balance}
\begin{document}
\title{Neural Parameter-varying Data-enabled Predictive Control of Cold Atmospheric Pressure Plasma Jets}

\author{Pegah GhafGhanbari, Mircea Lazar, Javad Mohammadpour Velni
\thanks{Manuscript created on February 19, 2025; This work was supported by the US National Science Foundation (NSF) under awards CMMI-2302219 and CNS-2302215.

Pegah GhafGhanbari and Javad M. Velni are with the Department of Mechanical Engineering, Clemson University, Clemson, SC 29631 USA. (email: pghafgh@clemson.edu; javadm@clemson.edu)

Mircea Lazar is with the Department of Electrical Engineering, Eindhoven University of Technology, Eindhoven, The Netherlands. (email: m.lazar@tue.nl)}}


\maketitle

\begin{abstract}
Cold Atmospheric Pressure Plasma Jets (APPJs) show significant potential for biomedical applications, but their inherent complexity, characterized by nonlinear dynamics and strong sensitivity to operating conditions like tip-to-surface distance, presents challenges for real-time control. This paper introduces the Neural Parameter-varying Data-enabled Predictive Control (NPV-DeePC) framework to address these issues. By integrating hypernetworks into the neural DeePC paradigm, NPV-DeePC adaptively captures system nonlinearities and parameter variations, \textcolor{black}{dynamically adjusts the neural network's learned representation of the system, enabling accurate multi-step trajectory prediction and control. Simulation studies on surface temperature tracking and thermal dose delivery demonstrate that NPV-DeePC achieves higher accuracy and adaptability than existing controllers. Moreover, its computational efficiency supports real-time implementation, making it a practical approach for precise APPJ control and a generalizable solution for other nonlinear, parameter-varying systems.}
\end{abstract}


\section{Introduction}
\IEEEPARstart{C}{old} Atmospheric Pressure Plasma Jets (APPJs) are emerging as a transformative technology in biomedical applications, offering unique capabilities in wound healing \cite{schmidt2017cold}, targeted cancer therapy \cite{metelmann2015head}, skin cosmetology and dermatology \cite{busco2020emerging}, and bio-decontamination \cite{nicol2020antibacterial}. Their ability to generate \textcolor{black}{reactive oxygen and nitrogen species (RONS)} at low temperatures enables therapeutic efficacy while minimizing thermal damage, making them especially suited for non-invasive and sensitive procedures.

Despite their promise, the practical deployment of APPJs in biomedical settings relies on effective real-time control strategies capable of ensuring precise and reliable operation. APPJs exhibit highly nonlinear dynamics and are sensitive to changes in operating conditions, particularly during therapeutic applications involving handheld devices. Variations in tip-to-surface distance over irregular surfaces cause rapid plasma fluctuations, impacting the delivery of reactive species and risking suboptimal treatment or thermal damage \cite{gidon2016model}. \textcolor{black}{These challenges highlight the need for predictive modeling approaches that can represent APPJ dynamics and support the development of adaptive control strategies.}

\textcolor{black}{Balancing model accuracy with computational feasibility presents a central challenge in APPJ control.} High-fidelity models, such as \cite{van2009plasma}, capture complex APPJ dynamics but are computationally too intensive for real-time control, while simplified models often fail to represent nonlinear behaviors adequately. Recent advancements have shifted toward data-driven approaches, which offer flexibility for managing nonlinear dynamics within Model Predictive Control (MPC) frameworks. For instance, a stochastic MPC incorporating a Linear Time-Invariant (LTI) model derived through subspace identification, together with Gaussian Process (GP) regression to compensate for plant–model mismatches, was proposed in \cite{bonzanini2021learning}.
\textcolor{black}{Subsequent approaches improved accuracy and robustness by combining an LTI model with Bayesian Neural Networks (BNNs) for uncertainty modeling in scenario-based MPC framework \cite{bao2022learning} and by adopting Linear-Parameter-Varying (LPV) representations derived from Artificial Neural Networks (ANNs) to better capture system nonlinearities \cite{ghafghanbari2025learning}. }

\textcolor{black}{All these methods fall under indirect data-driven control approaches and require an accurate system identification phase before controller design. However, system identification can be resource-intensive and may yield models that are either overly complex or misaligned with control objectives due to a focus on data-fitting accuracy. While control-oriented identification can be designed to yield models that balance fidelity with closed-loop performance requirements \cite{gevers2005identification}, achieving this alignment demands careful consideration of the application’s needs. To overcome these limitations, direct data-driven control approaches like Data-enabled Predictive Control (DeePC) \cite{coulson2019data} have gained significant attention. Building on behavioral systems theory and Willems' fundamental lemma \cite{willems2005note} for LTI systems, DeePC predicts future outputs directly from structured input-output data, therby bypassing the need for explicit system models.} 

Extensions have been developed to handle stochastic and nonlinear systems including regularization \cite{coulson2019data,elokda2021data}, feedback linearization \cite{alsalti2023data}, Koopman operator-based lifting \cite{korda2018linear}, basis function expansions \cite{lazar2024basis}, \textcolor{black}{and LPV formulations for systems with affine scheduling dependencies \cite{verhoek2023linear}. More recently, Neural DeePC has been proposed, where basis functions derived from the final hidden layer of a deep neural network (DNN) enrich the data representation, enabling DeePC to capture complex nonlinear dynamics more effectively \cite{lazar2024neural}.}

\textcolor{black}{Building on the foundation of Neural DeePC \cite{lazar2024neural}, this work introduces the Neural Parameter-Varying Data-enabled Predictive Control (NPV-DeePC) framework, a novel approach for controlling the evolving nonlinear dynamics of APPJs. Like Neural DeePC, NPV-DeePC uses a deep neural network, trained offline, to represent system dynamics and predict multi-step behavior for improved control accuracy. Unlike Neural DeePC, however, NPV-DeePC employs hypernetworks to form a HyperDNN \cite{chauhan2024brief}, which dynamically adjusts the DNN’s internal parameters in response to changing conditions. This enables NPV-DeePC to represent a continuous family of system models, maintaining high-precision control across a wide range of operating conditions with minimal online computational overhead. 
The second contribution is a conditioning-aware training strategy that augments the HyperDNN’s objective with a regularization term. This encourages well-structured and informative data representations, helping maintain the numerical properties required for reliable control synthesis. Overall, these contributions improve adaptability and robustness to noise, positioning NPV-DeePC as a practical solution for parameter-varying systems, including APPJs in dynamic biomedical contexts.}

This paper is organized as follows. Section II describes the problem and defines the control objectives for APPJs. Section III introduces the NPV-DeePC framework, detailing the DNN architecture and its integration into the DeePC approach. Section IV presents simulation results validating the proposed method, and Section V concludes with a summary of key findings and directions for future work.

\subsection{Notations}
Let $\mathbb{N}$ and $\mathbb{R}$ denote the sets of natural and real numbers, $\mathbb{N}_{\geq 1}$ the set of natural numbers excluding zero, $\mathbb{R}_{\geq 0}$ the set of non-negative real numbers, and $\mathbb{R}^n$ the set of real vectors of dimension $n \times 1$. \textcolor{black}{For any finite number $m \in \mathbb{N}_{\geq 1}$ of vectors or functions $\{v_1, \hdots, v_{m} \}$, we introduce the vertical stacking as $\text{col}(v_1, \hdots, v_{m})$. The Moore-Penrose pseudo-inverse of a matrix is denoted by $^\dagger$. For a square, symmetric matrix $\mathsf{M}$, $\mathsf{M}\succeq 0$ indicates that $\mathsf{M}$ is positive semi-definite, while $\mathsf{M} \succ 0$ signifies that $\mathsf{M}$ is positive definite. The Frobenius norm of a matrix $\mathsf{M}\in \mathbb{R}^{m\times n}$ is denoted by $\|\mathsf{M}\|_\mathfrak{F} = \left( \sum_{i=1}^m \sum_{j=1}^n |\mathsf{M}_{ij}|^2 \right)^{1/2}$. The weighted $l_2$-norm of a vector $v$ is defined as $\|v\|_\mathsf{M} = (v^\mathsf{T} \mathsf{M} v)^{1/2}$, where $\mathsf{M}$ is a positive (semi-) definite matrix.} \textcolor{black}{For the sequence $\left\{v(1), v(2), \hdots, v(L)   \right \}$  with elements $v(i) \in \mathbb{R}^{n_v}$, the Hankel matrix $\mathcal{H}_{D}\in \mathbb{R}^{{n_v D}\times(L-D+1)}$ of depth $D$, for some $D \leq L$, is defined as}
\begin{equation}\textcolor{black}{
    \mathcal{H}_{D}(v)=
    \begin{bmatrix}
        v(1) & v(2) & \hdots & v(L-D+1)\\
        v(2) & v(3) & \hdots & v(L-D+2)\\
        \vdots & \vdots & \ddots & \vdots\\
        v(D) & v(D+1) & \hdots & v(L)\\
    \end{bmatrix}.}
\end{equation}

\section{Problem Description}
This section describes the structure of APPJs, the challenges in achieving precise control, and the objectives guiding the control design.

\subsection{Structure of Atmospheric Pressure Plasma Jets}
\begin{figure}
\centering
\includegraphics[width=4.6cm]{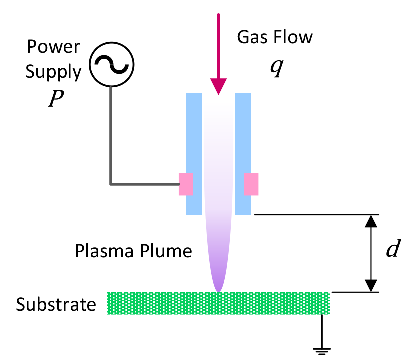}
\caption{Schematic of atmospheric pressure plasma jets.}
\label{fig:plasma-jet}
\end{figure}

APPJs, particularly in biomedical applications, are compact, hand-held devices that generate plasma by applying an electric field to a carrier gas, typically a noble gas like helium (He) or argon (Ar). This process creates a plasma plume that extends several centimeters beyond the device nozzle, as illustrated in Fig. \ref{fig:plasma-jet} \cite{laroussi2005room}.

At the core of an APPJ is a dielectric capillary tube through which the carrier gas flows at rates of a few slms \textcolor{black}{(standard liters per minute)}. \textcolor{black}{An electrode system, either surrounding or embedded within the tube, applies pulsed or sinusoidal voltages in the range of hundreds of Hz to several MHz. This ionizes and excites the gas molecules, initiating a plasma discharge that extends as a plume beyond the nozzle \cite{gidon2016model}. The discharge produces RONS, ultraviolet (UV) and visible-range photons, electric fields, and localized thermal effects \cite{laroussi2004evaluation}.}

The products of plasma discharge are central to many biomedical applications. RONS, for instance, drive key chemical reactions and play a critical role in cancer therapy. They selectively disrupt cancer cell functions, inducing apoptosis while sparing healthy cells \cite{dubey2022emerging}. UV radiation emitted during plasma discharge serves as an effective disinfectant, inducing thymine dimerization in bacterial DNA and disrupting replication. Although APPJs operate at near-room temperatures, their localized thermal effects are still significant in applications such as thermal coagulation \cite{robotis2003argon}, as well as inducing hyperthermia and thermal stress in cellular metabolism.

The characteristics of the plasma plume depend on several factors, including the electrode geometry, gas flow rate and composition, applied voltage, and ambient conditions such as temperature and humidity. With an optimal configuration and control strategy, the plasma plume can effectively treat complex and uneven surfaces, offering flexibility for diverse biomedical treatments.

\subsection{Control Challenges for APPJs}

\textcolor{black}{The effectiveness of model-based control approaches, such as MPC, depends critically on the availability of a precise yet computationally tractable dynamic model. For APPJs, developing such a model presents a significant challenge due to their inherently complex, nonlinear dynamics spanning multiple temporal and spatial scales \cite{haas1973plasma}. These nonlinearities arise from several interrelated phenomena. Gas flow transitions between laminar and turbulent regimes, governed by the Reynolds number, induce hydrodynamic fluctuations that destabilize plasma. In addition, the plasma’s electrodynamic response exhibits nonlinear behavior, influenced by gas composition, flow rate, and applied voltage, leading to abrupt mode transitions \cite{iseni2014atmospheric}. These difficulties are further compounded by the strong coupling among heat transfer, electric fields, and RONS generation.} 

\textcolor{black}{The practical deployment of APPJs, particularly in plasma biomedicine, introduces further control challenges due to their hand-held operation. Variations in the tip-to-surface distance during use alter plasma properties, such as plume length, RONS generation, and heat transfer, which directly affect treatment efficacy and safety. Such variations exacerbate the difficulties posed by nonlinear dynamics and parameter uncertainty, necessitating effective modeling approaches and adaptive control strategies to ensure reliable performance across diverse operating conditions \cite{gidon2016model}.}

\textcolor{black}{Traditional first-principles models, typically formulated as multidimensional partial differential equations (PDEs), are computationally prohibitive for real-time control \cite{breden2012self}. Reduced-order formulations, such as one-dimensional PDE models, offer greater tractability for control-oriented modeling\cite{gidon2017effective}, but their accuracy relies on precise parameter identification, which is often difficult to extract from experimental data. Furthermore, due to the assumptions made in deriving these models, they may omit dynamics that are critical for certain control tasks. To overcome these limitations, data-driven methods have emerged as a compelling alternative, offering high-fidelity approximations of APPJ dynamics with low computational cost \cite{gidon2021data}.}

\subsection{Control Objectives for APPJs}
\textcolor{black}{The dynamic behavior of APPJ can be described by the following discrete-time nonlinear parameter-varying model:
\begin{align}\label{eq:dyn_mdl}
    \nonumber x(k+1) &= f(x(k), u(k), p(k)), \\
    y(k) &= h(x(k), u(k)),
\end{align} 
where \( x \in \mathbb{R}^{n_x} \), \( u \in \mathbb{R}^{n_u} \), and \( p \in \mathbb{R}^{n_p} \) represent the state, input, and time-varying physical parameter vectors. The functions \( f: \mathbb{R}^{n_x} \times \mathbb{R}^{n_u} \times \mathbb{R}^{n_p} \mapsto \mathbb{R}^{n_x} \) and \( h: \mathbb{R}^{n_x} \times \mathbb{R}^{n_u} \mapsto \mathbb{R}^{n_y} \) describe the system dynamics and measurement process.}

For APPJs, the key control variables are applied power ($P$) and gas flow rate ($q$), forming the input vector $u=[P~~q]^\mathsf{T}$. The measurable outputs, represented by $y=[T_s~~T_g]^\mathsf{T}$, include substrate temperature ($T_s$) and gas temperature ($T_g$). The control objectives are primarily focused on ensuring effective plasma delivery while maintaining substrate safety and comfort. Cumulative Equivalent Minutes at $43~^{\circ}\text{C}$ (CEM) is a widely used metric to quantify thermal dose in hyperthermia therapies \cite{dewhirst2003basic}, which is defined as:
\begin{align}   \text{CEM}(k+1)&=\text{CEM}(k)+\mu^{(43-T_{s}(k))}\delta t,\\
    \nonumber \mu &=
    \begin{cases}
        0.5,& \text{if } T_s\geq 35~^{\circ}\text{C}\\
        0,              & \text{otherwise},
    \end{cases}
\end{align}

Here, $\delta t$ is the time increment, and the constant $\mu$ accounts for the thermal-stress response of the substrate. Achieving and maintaining the target CEM at the desired location is critical to balancing efficacy and tissue preservation, as deviations can lead to suboptimal treatment or irreversible damage to the substrate. Precise regulation of surface temperature is another control objective that optimizes system performance while ensuring substrate integrity and user comfort. Additionally, the stability of the plasma jet’s discharge regime must be maintained by constraining gas temperature, electric power, and current within appropriate ranges to avoid phenomena such as arcing or undesired mode transitions.

These objectives are further complicated by environmental variations, particularly fluctuations in the tip-to-substrate distance, which directly affect surface temperature, CEM, and plasma stability. \textcolor{black}{To capture this variability, we characterize the tip-to-surface distance as the system’s time-varying parameter $p$ to enable control adaptation and consistent performance across changing operating conditions.}

\color{black}
The control design objective is formulated as a constrained finite-horizon optimization problem to minimize a cost function $J: \mathbb{R}^{n_u} \times \mathbb{R}^{n_y} \mapsto \mathbb{R}_{\ge 0}$ over the predicted input and output vectors. At each time step $k$, the goal is to find an admissible input vector $\mathbf{u}(k):= \text{col}(u(k),\hdots,u(k+N-1)) \subset \mathbb{U}\subseteq\mathbb{R}^{n_u}$ over the \textit{prediction horizon}, $N\in\mathbb{N}_{\ge 1}$, such that the resulting output vector $\mathbf{y}(k):=\text{col}(y(k),\hdots,y(k+N-1))$ remains within the constraint set $\mathbb{Y}\subseteq\mathbb{R}^{n_y} $, while minimizing $J(\mathbf{u}(k),\mathbf{y}(k))$. The sets $\mathbb{U}$ and $\mathbb{Y}$ are assumed to be convex and compact, with $\mathbb{Y}$ containing the reference $r(k) \in \mathbb{R}^{n_y}$ in its interior. The physical parameter, $p(k)$, is updated at each time step but held constant over the prediction horizon. This is formulated as:
\begin{subequations}\label{eq:GenOpt}
\begin{flalign} 
    \min_{\mathbf{u}(k),\mathbf{y}(k)}& ~  J(\mathbf{u}(k),\mathbf{y}(k)), && \\ 
    \text{s.t.} \quad 
    & x(k+i+1) = f(x(k+i),u(k+i),p(k)), && \label{eq:DynConst}\\
    & y(k+i) = h(x(k+i),u(k+i)), && \label{eq:MeasConst}\\
    \nonumber&(u(k+i),y(k+i)) \in \mathbb{U} \times \mathbb{Y}, ~ i \in \{0,\hdots,N-1 \}
\end{flalign}
\end{subequations}
where $x(k)$ at $i=0$ is the measurement/estimation of states at time step $k$. The cost function, comprising stage and terminal costs, is defined as:
\begin{align} 
    \nonumber  &J(\mathbf{u}(k),\mathbf{y}(k))  := \sum_{i=0}^{N-1} \ell_s(y(k+i),u(k+i))+\ell_t(y(k+N)),\\
    \nonumber
     &\ell_s(\text{\small $y(k+i),u(k+i)$}):= \| y(k+i) - r(k) \|_{\mathsf{Q}}^2+ \|  \Delta u(k+i)\|_{\mathsf{R}}^2,\\
      &\ell_t(\text{\small $y(k+N)$}):= \| y(k+N) - r(k) \|_\mathsf{P}^2,
\end{align}
where $\Delta u(k+i) := u(k+i) - u(k+i-1)$ is the input increment. For $i=0$, $u(k-1)$ denotes the most recent control input applied to the system at the previous time step. The weighting matrices  $\mathsf{P} \succeq 0$, $\mathsf{Q} \succeq 0$, and $\mathsf{R} \succ 0$ tune the trade-off between reference tracking accuracy and smoothness of the control signal.

In this study, we propose a hybrid data-driven approach to formulate the dynamic constraints \eqref{eq:DynConst}–\eqref{eq:MeasConst} for controlling APPJs. The method combines indirect and direct data-driven control: a hypernetwork-based architecture (HyperDNN) modulates a neural network to predict multi-step-ahead dynamics under varying conditions, while DeePC exploits this model for real-time control. By embedding nonlinear behavior into a compact, adaptive representation, the HyperDNN enables the DeePC framework to efficiently compute optimal control actions under shifting dynamics. The detailed architecture and implementation are presented in the next section.
\color{black}

\section{Hybrid Data-driven Control Approach}
In this section, we introduce the Neural Parameter-Varying Data-Enabled Predictive Control (NPV-DeePC) framework, developed to address the nonlinear and parameter-varying dynamics of APPJs. \textcolor{black}{We first outline the mathematical preliminaries underlying the proposed strategy, then present the deep neural network architecture designed to capture system dynamics, and finally, describe the NPV-DeePC algorithm, which embeds this representation into a predictive control framework tailored to APPJs’ complex behavior.}

\subsection{Preliminaries}
Many \textit{indirect} data-driven predictive control approaches utilize multi-step predictions, typically employing \textcolor{black}{nonlinear autoregressive exogenous (NARX)} models, built from historical input-output data collected from the plant \cite{lazar2023nonlinear}.
The objective is to predict the future outputs of the system, $\mathbf{y}(k)$, given the past input-output trajectories $\{\mathbf{u}_{\text{ini}}(k),\mathbf{y}_{\text{ini}} (k)\}$ alongside the future inputs $\mathbf{u}(k)$ \cite{lazar2024basis}. In other words:
\begin{align}\label{eq:NARX}
        &\mathbf{y}(k) = \mathcal{F}(\mathbf{u}_{\text{ini}}(k),\mathbf{y}_{\text{ini}}(k),\mathbf{u}(k)), \\
        &\nonumber\mathbf{u}_{\text{ini}}(k):=\text{col} (u(k-T_{\text{ini}}),\hdots,u(k-1))\in \mathbb{R}^{n_u T_{\text{ini}}},\\
        &\nonumber\mathbf{y}_{\text{ini}}(k):=\text{col} (y(k-T_{\text{ini}}),\hdots,y(k-1))\in \mathbb{R}^{n_y T_{\text{ini}}},
\end{align}
where $T_{\text{ini}} \in \mathbb{N}_{\ge1}$ is the \textit{past} \textit{horizons}, and $\mathcal{F} := \text{col}(F_1, \hdots, F_N)$ represents a parameterized map, typically defined using neural networks. Each element $F_i$ in $\mathcal{F}$ is a multi-input-multi-output (MIMO) predictor and is formed by aggregating several multi-input-single-output (MISO) predictors, such that $F_i = \text{col}(F_{i,1}, \hdots, F_{i,n_y})$, where each $F_{i,j}$ predicts the $j$-th element of the output at time step $i$.

In \textit{direct} data-driven predictive control, behavioral system theory and Willems' fundamental lemma \cite{willems2005note} are central to predicting future system behavior based on observed input-output trajectories. A key requirement is the persistence of excitation, which guarantees that the collected data are sufficiently rich to represent the system dynamics.
\begin{definition} (Persistence of Excitation)
    The data sequence $\{v(k) \}_{k=1}^{L}$ with $v(k)\in \mathbb{R}^{n_v}$ is called persistently exciting of order $D$ if the Hankel matrix $\mathcal{H}_D(v)$ has full row rank, i.e., 
    \begin{equation}
        \text{rank}(\mathcal{H}_{D}(v)) = n_v D.
    \end{equation}
\end{definition}
This condition imposes a lower bound on the trajectory length, namely $L \ge (n_v + 1) D - 1$ \cite{markovsky2021behavioral}, and plays a crucial role in system identification and input design.

The following lemma, from behavioral systems theory, links controllability, persistence of excitation, and the column span of the Hankel matrix to represent all possible trajectories of an LTI system. 
\vspace{6pt}
\begin{lemma} \label{lm:Willems}
(Willems' Fundamental Lemma \cite{willems2005note}) 
\textcolor{black}{Consider an LTI system of order $n$. Suppose that following conditions hold:
\begin{enumerate}
    \item $\{\hat{u}(k), \hat{y}(k)\}_{k=1}^{L}$ is a trajectory of the system,
    \item The system is controllable, and
    \item The input $\hat{u}$ is persistently exciting of order $D + n$.
\end{enumerate}
Then, any $D$-samples long trajectory $\{(\tilde{u}(k), \tilde{y}(k))\}_{k=1}^{D}$ of the system can be represented as a linear combination of the columns of the Hankel matrices constructed from the original trajectory. That is, it is also a trajectory of the system if and only if there exists a vector $\mathbf{g} \in \mathbb{R}^{L - D + 1}$ such that:
\begin{equation}
    \begin{bmatrix}
        \mathcal{H}_D(\hat{u}) \\
        \mathcal{H}_D(\hat{y})
    \end{bmatrix}
    \mathbf{g} = 
    \begin{bmatrix}
        \tilde{u} \\
        \tilde{y}
    \end{bmatrix}.
\end{equation}}
\end{lemma}

\vspace{-18pt}
This states that the subspace spanned by the columns of the Hankel matrix corresponds exactly to the subspace of possible trajectories of the system. Building upon Willems' fundamental lemma, the DeePC approach reformulates the optimal control problem \eqref{eq:GenOpt} as
\begin{subequations}\label{eq:DeePC}
\begin{align}  
    \min_{\Xi}~ &J(\mathbf{u}(k),\mathbf{y}(k))+\lambda_{\varrho} \| \boldsymbol{\varrho}(k) \|^2+ \lambda_g \ell_g (\mathbf{g}(k)), \\
    \text{s.t.} \quad 
        &\begin{bmatrix}
        \mathcal{U}_p\\
        \mathcal{Y}_p\\
        \mathcal{U}_f\\
        \mathcal{Y}_f
    \end{bmatrix}
    \mathbf{g}(k) = 
    \begin{bmatrix}
        \mathbf{u}_{\text{ini}}(k)\\
        \mathbf{y}_{\text{ini}}(k)+ \boldsymbol{\varrho}(k)\\
        \mathbf{u}(k)\\
        \mathbf{y}(k)
    \end{bmatrix}, \label{eq:WillConst}\\
    & (\mathbf{u}(k), \mathbf{y}(k))\in \mathbb{U}^N\times \mathbb{Y}^N,
\end{align}
\end{subequations}
\textcolor{black}{where $\Xi := \text{col}(\mathbf{u}(k), \mathbf{y}(k), \mathbf{g}(k), \boldsymbol{\varrho}(k))$ denotes the collection of all optimization variables.}
The left-hand side of \eqref{eq:WillConst} is constructed by partitioning the Hankel matrices \textcolor{black}{built from previously recorded input-output trajectories of length $L$, $\{\hat{u}(k), \hat{y}(k)\}_{k = 1}^{L}$, as}
\begin{equation}
    \begin{bmatrix}
        \mathcal{U}_p\\
        \hline
        \mathcal{U}_f
    \end{bmatrix} =
    \mathcal{H}_{T_{\text{ini}}+N}(\hat{u}),\\
    \quad
    \begin{bmatrix}
        \mathcal{Y}_p\\
        \hline
        \mathcal{Y}_f
    \end{bmatrix} \sim
    \mathcal{H}_{T_{\text{ini}}+N}(\hat{y}),  
\end{equation}
where the subscripts ``\(p\)'' and ``\(f\)'' denote the past and future horizons, respectively.

\textcolor{black}{In the DeePC formulation \eqref{eq:DeePC}, two regularization terms, $\| \boldsymbol{\varrho}(k) \|^2$ and $\ell_g(\mathbf{g}(k))$, are introduced to enhance robustness and ensure feasibility in the presence of nonlinearities and process/measurement noise. Such imperfections violate Willems’ Fundamental Lemma, which assumes persistently exciting, noise-free data from an LTI system.}

\textcolor{black}{In practice, noise and nonlinear behavior render the Hankel matrix full-rank, so observed trajectories deviate from the ideal subspace spanned by system responses. To address this, the slack variable $\boldsymbol{\varrho}(k)$ is introduced in the initial trajectory constraint \eqref{eq:WillConst} to account for deviations due to imperfect data. A large penalty on $\boldsymbol{\varrho}(k)$ ensures that this variable remains small unless necessary, thereby preserving feasibility while maintaining fidelity to the measured data.}

\textcolor{black}{The second regularization term $\ell_g(\mathbf{g}(k))$ mitigates overfitting to noisy data and improves numerical conditioning. Standard norm-based approaches include $\ell_1$-norm \cite{coulson2019data}, $\ell_2$-norm \cite{xue2021data}, squared $\ell_2$-norm \cite{berberich2020data}, and generalized $p$-norm \cite{coulson2021distributionally}. While effective, such penalties act on the entire solution vector $\mathbf{g}(k)$ and may bias the predicted trajectories. To overcome this, a projection-based regularization was proposed in \cite{dorfler2022bridging}:}
\begin{equation} \label{eq:regularization}
    \ell_g (\mathbf{g}(k)):=\| (I-\Pi)\mathbf{g}(k) \|^2, \quad \Pi=
    \begin{bmatrix}
        \mathcal{U}_p\\
        \mathcal{Y}_p\\
        \mathcal{U}_f
    \end{bmatrix}^\dagger
    \begin{bmatrix}
        \mathcal{U}_p\\
        \mathcal{Y}_p\\
        \mathcal{U}_f
    \end{bmatrix}.
\end{equation}

\textcolor{black}{Here, $I$ denotes the identity matrix of appropriate dimension, and $(I-\Pi)$ is the orthogonal projector onto the kernel of the first three block-constraint equations in \eqref{eq:WillConst}. Unlike norm-based regularizers, it penalizes only the homogeneous component of the solution, leaving the predicted inputs and outputs unaffected and thus preserving trajectory consistency.}

\textcolor{black}{Although regularization improves DeePC’s ability to handle nonlinear systems, the method still relies on linear combinations of measured trajectories to approximate nonlinear behavior. Enlarging the Hankel matrix can enhance this approximation but at the cost of prohibitive computational complexity, limiting real-time applicability. To overcome this, a neural network can be integrated into the DeePC framework to learn the system’s nonlinear dynamics directly. This shifts the modeling effort from the Hankel-based trajectory space to an offline–trained nonlinear representation, achieving higher predictive accuracy with improved computational efficiency.}

\subsection{Hyper Deep Neural Network Architecture}
\color{black}
The proposed hybrid control framework incorporates an offline training phase to capture APPJ dynamics under varying operating conditions. Building on the neural DeePC framework, we employ a HyperDNN that dynamically generates network parameters for multi-step output prediction. This design improves representation capacity while keeping computational overhead during inference minimal.

\begin{figure}
\centering
\includegraphics[width=1\columnwidth]{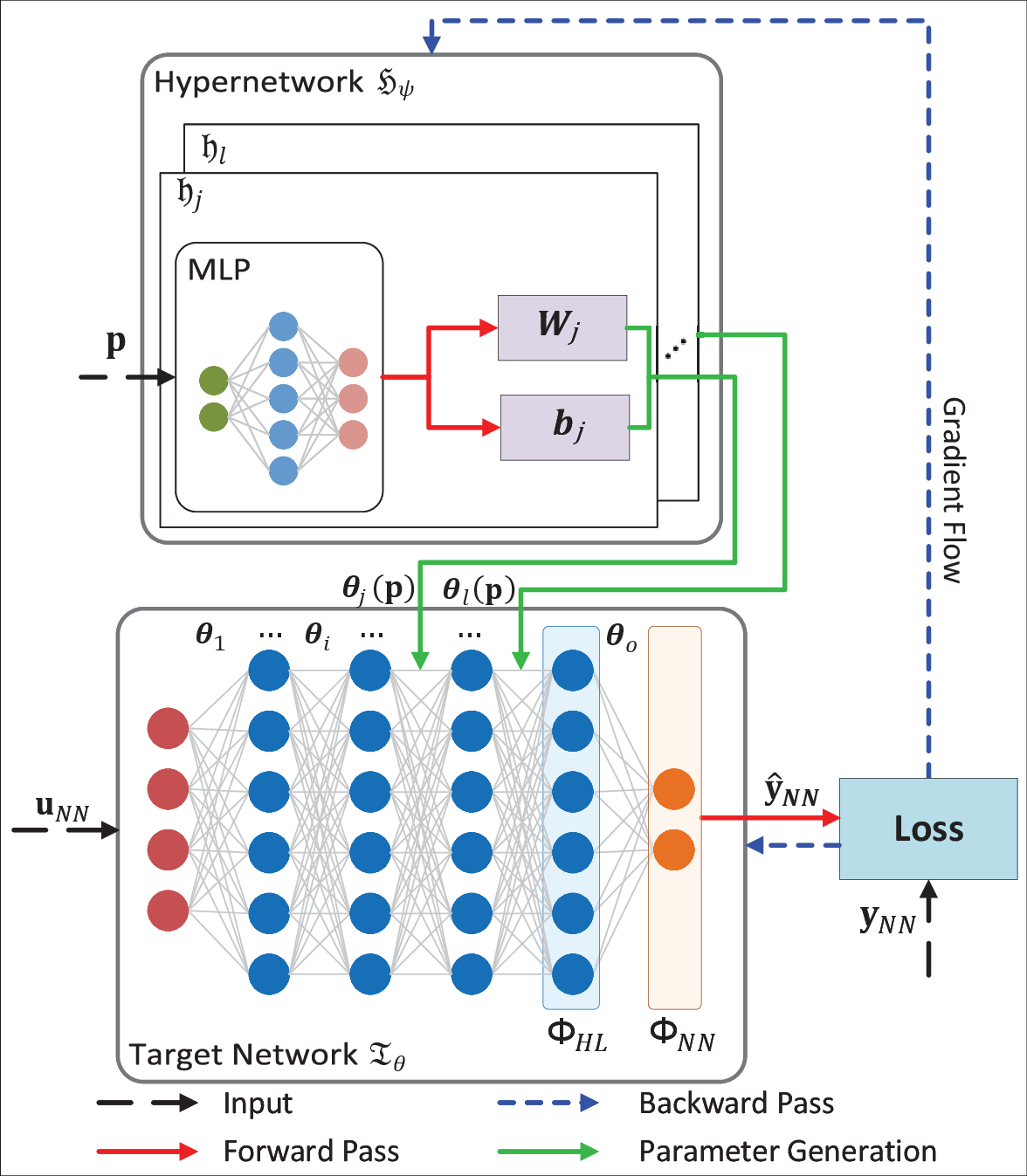}
\caption{\textcolor{black}{Overview of the proposed HyperDNN architecture. The target network $\mathfrak{T}_\theta$ predicts the output $\hat{\mathbf{y}}_{NN}$ from the input $\mathbf{u}_{NN}$, constructing the neural space $\upphi_{HL}$. Its parameters include fixed weights and biases for the first $m$ hidden layers, capturing general features, and dynamic parameters for the subsequent layers, computed by the hypernetwork $\mathfrak{H}_\psi$. The hypernetwork consists of subnetworks $\mathfrak{h}_j$, each generating weights and biases for the corresponding $j$-th dynamic hidden layer of $\mathfrak{T}_\theta$ based on the contextual input $\mathbf{p}$, enabling adaptation to varying APPJ operating conditions.}}
\label{fig:DNN}
\end{figure}

As illustrated in Fig. \ref{fig:DNN}, the HyperDNN consists of two main components: a target network, $\mathfrak{T}_\theta$, and a hypernetwork, $\mathfrak{H}_\psi$, where $\theta$ and $\psi$ denote the learnable parameters of each network, respectively. The target network performs the multi-step prediction task, while the hypernetwork dynamically generates the parameters of $\mathfrak{T}_\theta$ according to the system’s operating context \cite{ha2017hypernetworks}. To capture variations in operating conditions, $\mathfrak{H}_\psi$ processes a history of the physical parameter vector $p$ as contextual input, thereby adapting the target network online. The contextual input at time step $k$ is defined as
$$
\mathbf{p}(k) := \text{col}(p(k - T_{\text{ini}}), \dots, p(k - 1)) \in \mathbb{R}^{n_p T_{\text{ini}}}.
$$

The target network $\mathfrak{T}_\theta$ has $l$ layers in total. To balance computational efficiency and adaptability, the first $m$ layers are fixed, while the remaining $l-m$ layers are dynamically adapted by the hypernetwork $\mathfrak{H}_\psi$. Specifically, the parameters of the $m$ fixed layers are 
$$
\boldsymbol{\theta}_\mathfrak{T} := \{ \theta_i \}_{i=1}^m, \quad \theta_i = [W_i ~~ b_i],
$$
which are directly learned during training and capture general features that remain valid across a wide range of operating conditions. The value of $m$ is chosen empirically based on validation results to balance accuracy and training stability.

The hypernetwork $\mathfrak{H}_\psi$ generates the parameters for the remaining layers,
$$
\boldsymbol{\theta}_\mathfrak{H} := \{ \theta_j \}_{j=m+1}^l = \mathfrak{H}_\psi(\mathbf{p}(k)), \quad \theta_j = [W_j(\mathbf{p}) ~~ b_j(\mathbf{p})],
$$
where $\mathfrak{H}_\psi=\{\mathfrak{h}_j\}_{j=m+1}^l$ consists of subnetworks $\mathfrak{h}_j$, each producing the weights and biases for the $j$-th target layer. Subnetwork $\mathfrak{h}_j$ is parameterized by $\psi_j := \{ \psi_j^\iota \}_{\iota=1}^{\lambda_j}$, with $\lambda_j$ denoting its number of layers.

Finally, the complete parameter set of the target network is
$$
\boldsymbol{\theta} = \{\boldsymbol{\theta}_\mathfrak{T}, \boldsymbol{\theta}_\mathfrak{H}, \theta_o\}, \quad \theta_o = [W_o ~~ b_o],
$$
where $\theta_o$ denotes the fixed parameters of the output layer.

Both $\mathfrak{T}_\theta$ and $\mathfrak{H}_\psi$ are implemented as multilayer perceptrons (MLPs). Hidden layers use a nonlinear activation function $\sigma: \mathbb{R} \to \mathbb{R}$, applied element-wise to a vector $v \in \mathbb{R}^{n_v}$ as $\boldsymbol{\sigma}(v) := \text{col}(\sigma(v_1), \dots, \sigma(v_{n_v}))$. The output layers employ a linear activation function, $\sigma_{\text{lin}}(v) := v$, to produce either the updated weights for the target network or the final predictions.

At each sampling instant $k$, the target network processes the sequential input vector
$$
\mathbf{u}_{NN}(k) := \text{col}(\mathbf{u}_{\text{ini}}(k), \mathbf{y}_{\text{ini}}(k), \mathbf{u}(k)) \in \mathbb{R}^{(n_u + n_y)T_{\text{ini}} + n_u N},
$$
comprising past trajectories and planned future inputs to predict the future outputs $\hat{\mathbf{y}}_{NN}(k) := \hat{\mathbf{y}}(k) \in \mathbb{R}^{n_y N}$ through the following recursive computation:
\begin{align}\label{eq:rec_NN}
    \nonumber &\mathbf{z}_1 := \boldsymbol{\sigma}(W_1 \mathbf{u}_{NN} + b_1),\\
    \nonumber &\mathbf{z}_i := \boldsymbol{\sigma}(W_i \mathbf{z}_{i-1}  + b_i),\qquad\qquad i=2,\hdots,m,\\
    \nonumber &\mathbf{z}_j := \boldsymbol{\sigma}(W_j(\mathbf{p}) \mathbf{z}_{j-1}  + b_j(\mathbf{p})),\quad j=m+1,\hdots,l,\\
    &\hat{\mathbf{y}}_{NN} := \boldsymbol{\sigma}_{\text{lin}}(W_o \mathbf{z}_{l} +b_o)=W_o \mathbf{z}_{l} + b_o.
\end{align} 

Here, $i$ indexes the fixed layers, whereas $j$ indexes the adaptive layers with parameters provided by $\mathfrak{H}_\psi$. For brevity, the dependency on $k$ is omitted.

The HyperDNN is trained end-to-end, optimizing all parameters simultaneously. In the forward pass, the target network computes $\hat{\mathbf{y}}_{NN}$ from the sequential input $\mathbf{u}_{NN}$, while the hypernetwork adapts its parameters based on the contextual input $\mathbf{p}$. The loss function then quantifies the error between the ground truth $\mathbf{y}$ and the predicted output $\hat{\mathbf{y}}_{NN}$. In the backward pass, gradients are computed and backpropagated through both the target network and the hypernetwork, thereby updating the fixed parameters $\boldsymbol{\theta}_\mathfrak{T}$, the output layer parameters $\theta_o$, and the hypernetwork parameters $\boldsymbol{\psi}:=\{ \psi_j\}_{j=m+1}^l$.

We define the \textit{neural space} as the output of the last hidden layer of the target network, $\mathbf{z}_l = \upphi_{HL}(\mathbf{u}_{NN}, \mathbf{p})$, obtained through recursive affine transformations and nonlinear activations applied to the sequential input $\mathbf{u}_{NN}$ and contextual parameter vector $\mathbf{p}$, as described in \eqref{eq:rec_NN}. This mapping is formally given by $\upphi_{HL}: \mathbb{R}^{(n_u + n_y)T_{\text{ini}} + n_u N} \times \mathbb{R}^{n_p T_{\text{ini}}} \to \mathbb{R}^{n_l}$, where $n_l$ is the number of neurons in the last hidden layer. The overall HyperDNN mapping is then 
\begin{align} \label{eq:neuralSpace}
    \nonumber \upphi_{NN}(\mathbf{u}_{NN},\mathbf{p}) &:=W_o \mathbf{z}_{l} + b_o \\
     &=: W_o \upphi_{HL} (\mathbf{u}_{NN},\mathbf{p})+ b_o,
\end{align}
where $\upphi_{NN}: \mathbb{R}^{(n_u + n_y)T_{\text{ini}} + n_u N} \times \mathbb{R}^{n_p T_{\text{ini}}} \to \mathbb{R}^{n_y N}$
produces the predicted output $\hat{\mathbf{y}}_{NN}(k)$. This formulation highlights that $\upphi_{NN}$ performs an affine transformation on the neural space features $\upphi_{HL}$. While not guaranteed in general, such features are very likely to form a linearly independent basis \cite{chen1995universal}, allowing the HyperDNN to represent the nonlinear dynamics of APPJs as a linear combination of these features in neural space. This linear structure mirrors Willems' fundamental lemma, which represents system behavior as linear combinations of input-output data, thereby enabling seamless integration of the HyperDNN into the DeePC framework for data-driven control.

\color{black}

\subsection{Neural Parameter-Varying DeePC}
\color{black}
Neural Parameter-Varying DeePC (NPV-DeePC) extends the DeePC framework by integrating the HyperDNN, yielding a learned dynamics model conditioned on the physical parameter sequence $\mathbf{p}$ in place of the Hankel-based prediction of standard DeePC. This enables accurate trajectory prediction and effective control of nonlinear systems like APPJs across varying operating conditions.

To incorporate this model, the mappings $\upphi_{NN}$ and $\upphi_{HL}$ are applied to the Hankel matrix built from $L$ historical samples, $\mathscr{H} := \text{col}(\mathcal{U}_p, \mathcal{Y}_p, \mathcal{U}_f) \in \mathbb{R}^{((n_u + n_y)T_{\text{ini}} + n_u N) \times K}$,
with \( K = L - T_{\text{ini}} - N + 1 \),  together with the physical parameter Hankel matrix, $\mathcal{P} = \mathcal{H}_{T_{\text{ini}}}(p)\in \mathbb{R}^{n_pT_\text{ini} \times K}$. All columns of these Hankel matrices, denoted by \( \mathscr{H}_{:j} \) and \( \mathcal{P}_{:j} \) for \( j\in\{1,\dots,K \}\), are projected using the mappings \( \upphi_{HL} \) and \( \upphi_{NN} \) to form the following lifted Hankel matrices:
\begin{subequations}  
\begin{align}\label{eq:transHank}
    \boldsymbol{\Upphi}_{HL} &:= \begin{bmatrix} \upphi_{HL}(\mathscr{H}_{:1}, \mathcal{P}_{:1}) & \dots & \upphi_{HL}(\mathscr{H}_{:K}, \mathcal{P}_{:K}) \end{bmatrix} \\
    \nonumber &\in \mathbb{R}^{n_l \times K}, \\
    \boldsymbol{\Upphi}_{NN} &:= \begin{bmatrix} \upphi_{NN}(\mathscr{H}_{:1}, \mathcal{P}_{:1}) & \dots & \upphi_{NN}(\mathscr{H}_{:K}, \mathcal{P}_{:K}) \end{bmatrix} \\
    \nonumber &\in \mathbb{R}^{n_y N \times K}.
\end{align}
\end{subequations}

\color{black}
\textcolor{black}{During offline training, the network parameters are optimized by minimizing a loss function composed of two terms: (i) the squared Frobenius norm of the error between the Hankel matrix of true future outputs, $\mathcal{Y}_f$, and the HyperDNN-predicted outputs, $\boldsymbol{\Upphi}_{NN}$; and (ii) a condition-number regularization term that enforces well-conditioned features in $\text{col}(\boldsymbol{\Upphi}_{HL}, \boldsymbol{1}^\mathsf{T})$. The optimization problem is formulated as}
\begin{align} \label{eq:NLSprob}     
    \nonumber (\theta_o^*,\boldsymbol{\theta}_{\mathfrak{T}}^*,\boldsymbol{\psi}^*) = \arg \min &\left\|\mathcal{Y}_f-\boldsymbol{\Upphi}_{NN}(\theta_o,\boldsymbol{\theta}_{\mathfrak{T}},\boldsymbol{\psi})\right\|_{\mathfrak{F}}^{2}\\
    &\textcolor{black}{+\lambda_\kappa \kappa 
    \left (\begin{bmatrix}
             \boldsymbol{\Upphi}_{HL}(\boldsymbol{\theta}_{\mathfrak{T}},\boldsymbol{\psi})\\
              \boldsymbol{1}^\mathsf{T}
        \end{bmatrix} \right ),}
\end{align}
\textcolor{black}{where $\kappa(\cdot) = \varsigma_{{\max}}/\varsigma_{\min}$ denotes the condition number with $\varsigma_{\max}$ and $\varsigma_{\min}$ being the largest and smallest singular values, and $\lambda_\kappa$ serves as a coefficient that modulates the relative contribution of the condition number penalty to the overall objective.}

\textcolor{black}{The inclusion of the regularization term is motivated by the persistence excitation (PE) condition from Willems’ lemma, which requires that $\text{col}(\boldsymbol{\Upphi}_{HL}, \boldsymbol{1}^\mathsf{T})$ has full row rank. This guarantees that the neural basis functions form a sufficiently rich representation to capture the underlying nonlinear dynamics. By explicitly penalizing the condition number, the training process not only encourages this completeness but also mitigates potential numerical instability in subsequent least-squares computations.}

\textcolor{black}{The joint optimization \eqref{eq:NLSprob} typically converges to a local minimum due to the non-convex nature of the problem, which may result in suboptimal output layer parameters for the learned feature representation.}
To further enhance prediction accuracy, the parameters of the output layer can be refined by solving the following least-squares optimization problem:
\begin{equation}\label{eq:LSprob}
    \theta_o^{LS} =
    \arg \min \left\|\mathcal{Y}_f-\theta_0
    \begin{bmatrix}
         \boldsymbol{\Upphi}_{HL}(\boldsymbol{\theta}_{\mathfrak{T}}^*,\boldsymbol{\psi}^*)\\
          \boldsymbol{1}^\mathsf{T}
    \end{bmatrix}
    \right\|_{\mathfrak{F}}^2,
\end{equation}
where $\theta_o^{LS} = [W_o^{LS} ~~ b_o^{LS}]$, \textcolor{black}{and $\boldsymbol{1}\in \mathbb{R}^{K}$ is the vector of ones. This refinement step optimizes the output layer parameters, $\theta_o^{LS}$, to achieve the optimal linear mapping from the optimized neural space features, \(\boldsymbol{\Upphi}_{HL}(\boldsymbol{\theta}_{\mathfrak{T}}^*,\boldsymbol{\psi}^*)\), to the target outputs, \(\mathcal{Y}_f\). Since (\ref{eq:LSprob}) is a convex problem, it ensures a prediction error at least as good as, or better than, that of (\ref{eq:NLSprob}), thereby enhancing accuracy without altering the other network parameters.}

\textcolor{black}{Using the optimal parameters, $(\theta_o^{LS}, \boldsymbol{\theta}_{\mathfrak{T}}^*, \boldsymbol{\psi}^*)$, the multi–step input–output predictor is defined as:}
\begin{equation}\label{eq:predNARX}
    \mathbf{y}^\text{NLS}(k) = \upphi_{NN}(\mathbf{u}_{NN}(k),\mathbf{p}(k),\boldsymbol{\theta}_{\mathfrak{T}}^*,\boldsymbol{\psi}^*, \theta_o^{LS}).
\end{equation}

Given the affine relationship of neural space features and based on Willems' fundamental lemma, the NPV-DeePC prediction belongs to the following set:
\begin{multline}\label{eq:predNPV}
    \mathbf{y}^\text{NPV-DeePC}(k)\in\\
    \Bigg\{ \mathcal{Y}_f\mathbf{g}(k) : 
    \begin{bmatrix}
         \boldsymbol{\Upphi}_{HL}\\
          \boldsymbol{1}^\mathsf{T}
    \end{bmatrix}
    \mathbf{g}(k) =
    \begin{bmatrix}
         \upphi_{HL}(\mathbf{u}_{NN}(k),\mathbf{p}(k))\\
          1
    \end{bmatrix}
    \Bigg\}.
\end{multline}

To simplify the notation, we define $\upphi_{HL}(k):=\upphi_{HL}(\mathbf{u}_{NN}(k),\mathbf{p}(k))$ to represent the vector of neural features computed at sample time $k$ given the neural network inputs $\mathbf{u}_{NN}(k)$ and the varying system parameters $\mathbf{p}(k)$. The dynamic constraints in \eqref{eq:DynConst} and \eqref{eq:MeasConst} can now be replaced by:
\begin{equation}\label{eq:NPV_behavior2}
     \begin{bmatrix}
        \boldsymbol{\Upphi}_{HL}\\
        \mathbf{1}^\mathsf{T}\\
        \mathcal{Y}_f
    \end{bmatrix}
    \mathbf{g}(k)=
    \begin{bmatrix}
         \upphi_{HL}(k)\\
          1\\
          \mathbf{y}(k)
    \end{bmatrix}.
\end{equation}

The following lemma establishes the conditions for equivalence between the NLS prediction and NPV-DeePC models.
\begin{lemma} \label{lemma:eqCond}
Consider the nonlinear least squares optimal prediction model \eqref{eq:predNARX} and the NPV-DeePC model \eqref{eq:predNPV}, both constructed using the same dataset. Let
\begin{enumerate}
    \item $\mathcal{E}:= \mathcal{Y}_f - 
    \theta_o^{LS} 
    \begin{bmatrix}
         \boldsymbol{\Upphi}_{HL}\\
          \boldsymbol{1}^\mathsf{T}
    \end{bmatrix}$ be the residual matrix of the \textcolor{black}{least-squares problem} \eqref{eq:LSprob},
    \item $
    \mathcal{S}_g := \left\{  
        \begin{bmatrix}
            \boldsymbol{\Upphi}_{HL}\\
            \boldsymbol{1}^\mathsf{T}
        \end{bmatrix}^\dagger
        \begin{bmatrix}
            \upphi_{HL}(k)\\
            1
        \end{bmatrix} 
        + \hat{\mathbf{g}} :
         \hat{\mathbf{g}} \in \text{null} \left( \begin{bmatrix}
            \boldsymbol{\Upphi}_{HL}\\
            \boldsymbol{1}^\mathsf{T}
        \end{bmatrix} \right)
    \right\}$ be a set of parameters $\mathbf{g}$,
    \item $\begin{bmatrix}
         \boldsymbol{\Upphi}_{HL}\\
          \boldsymbol{1}^\mathsf{T}
    \end{bmatrix}$ has full row rank.
\end{enumerate}
Then, the two models \eqref{eq:predNARX} and \eqref{eq:predNPV} are equivalent \textit{if and only if} $\mathcal{E}\hat{\mathbf{g}}= \mathbf{0}$ for all  $\hat{\mathbf{g}} \in \text{null}\left(\begin{bmatrix}
         \boldsymbol{\Upphi}_{HL}\\
          \boldsymbol{1}^\mathsf{T}
    \end{bmatrix}\right)$.
\end{lemma}

\renewcommand\qedsymbol{$\blacksquare$}
\begin{proof}
Proof of this lemma is an extension of that in \cite{lazar2024neural}. From \eqref{eq:predNPV}, it is evident that
\[
    \begin{bmatrix}
         \boldsymbol{\Upphi}_{HL}\\
          \boldsymbol{1}^\mathsf{T}
    \end{bmatrix}
    \mathbf{g}(k) =\\ 
    \begin{bmatrix}
         \upphi_{HL}(k)\\
          1
    \end{bmatrix},
\]
indicating that all $\mathbf{g}(k)$ satisfying this equation lie in the set $\mathcal{S}_g$. Consequently, the predicted outputs generated by NPV-DeePC satisfy:
\[
\begin{aligned}
&\mathbf{y}^\text{NPV-DeePC}(k)\in\\
&\left\{  
    \mathcal{Y}_f \Bigg(
        \begin{bmatrix}
            \boldsymbol{\Upphi}_{HL}\\
            \boldsymbol{1}^\mathsf{T}
        \end{bmatrix}^\dagger
        \begin{bmatrix}
            \upphi_{HL}(k)\\
            1
        \end{bmatrix} 
        + \hat{\mathbf{g}}\Bigg) :
         \hat{\mathbf{g}} \in \text{null} \left( \begin{bmatrix}
            \boldsymbol{\Upphi}_{HL}\\
            \boldsymbol{1}^\mathsf{T}
        \end{bmatrix} \right)
    \right\}.
    \end{aligned}
    \]
    Then, from the definition of the residual and considering that $\hat{\mathbf{g}}$ belongs to the null space of $\text{col}( \boldsymbol{\Upphi}_{HL},\boldsymbol{1}^\mathsf{T})$, we can write:
    \[
     \mathcal{Y}_f \hat{\mathbf{g}}=\mathcal{E} \hat{\mathbf{g}}+ \theta_o^{LS} 
    \begin{bmatrix}
         \boldsymbol{\Upphi}_{HL}\\
          \boldsymbol{1}^\mathsf{T}
    \end{bmatrix} \hat{\mathbf{g}}= \mathcal{E}\hat{\mathbf{g}}.
    \]
    Ideally, when $\mathcal{E}=\mathbf{0}$, we have $\theta_o^{LS} =\mathcal{Y}_f  
    \begin{bmatrix}
         \boldsymbol{\Upphi}_{HL}\\
          \boldsymbol{1}^\mathsf{T}
    \end{bmatrix}^\dagger$, thus
    \[
    \mathbf{y}^\text{NPV-DeePC}(k)=
     \mathcal{Y}_f 
        \begin{bmatrix}
            \boldsymbol{\Upphi}_{HL}\\
            \boldsymbol{1}^\mathsf{T}
        \end{bmatrix}^\dagger
        \begin{bmatrix}
            \upphi_{HL}(k)\\
            1
        \end{bmatrix} =\mathbf{y}^\text{NLS}(k),
    \]
    if and only if $\mathcal{E} \hat{\mathbf{g}}=\mathbf{0}$ for all $\hat{\mathbf{g}} \in \text{null} \left(\begin{bmatrix}
         \boldsymbol{\Upphi}_{HL}\\
          \boldsymbol{1}^\mathsf{T}
    \end{bmatrix}\right)$.
\end{proof}
Since the assumption $\mathcal{E} =\mathbf{0}$ does not generally hold in practice, a suitable regularization cost is required to penalize the resulting discrepancy. \textcolor{black}{One straightforward choice is 
\begin{align}\label{eq:NLS_regularization}
\ell_g(\mathbf{g})&:=\|\mathbf{g}(k)-\mathbf{g}^{\text{NLS}}(k) \|^2 ,\\
\nonumber \mathbf{g}^{\text{NLS}}(k) &:= 
\begin{bmatrix}
            \boldsymbol{\Upphi}_{HL}\\
            \boldsymbol{1}^\mathsf{T}
        \end{bmatrix}^\dagger
        \begin{bmatrix}
            \upphi_{HL}(k)\\
            1
        \end{bmatrix}.
\end{align}}

\textcolor{black}{This regularization, however, is highly nonlinear with a non–sparse structure, which results in significant computational complexity. As an alternative, the behavior constraint \eqref{eq:NPV_behavior2} can be reformulated by introducing $\hat{\mathbf{g}}$ as the optimization variable, defined as}
\begin{subequations}
\begin{align}
    &\begin{bmatrix}
        \boldsymbol{\Upphi}_{HL}\\
        \boldsymbol{1}^\mathsf{T}
    \end{bmatrix}
    \hat{\mathbf{g}}(k) =
    \begin{bmatrix}
        \mathbf{0}\\
        0
    \end{bmatrix}, \\
    &\mathcal{Y}_f\!\left(
        \begin{bmatrix}
            \boldsymbol{\Upphi}_{HL}\\
            \boldsymbol{1}^\mathsf{T}
        \end{bmatrix}^\dagger
        \begin{bmatrix}
            \upphi_{HL}(k)\\
            1
        \end{bmatrix} 
        + \hat{\mathbf{g}} \right) = \mathbf{y}(k).
\end{align}
\end{subequations}

This reformulation simplifies the regularization \eqref{eq:NLS_regularization} to an efficient term, $\ell_g = \| \hat{\mathbf{g}}(k) \|^2$ \cite{lazar2024neural}.
Assuming $\mathcal{Y}_f$ has full row rank, $\hat{\mathbf{g}}(k) \in \mathbb{R}^K$ can be reparameterized as $\tilde{\mathbf{g}}(k) := \mathcal{Y}_f \hat{\mathbf{g}}(k) \in \mathbb{R}^{n_y N}$, reducing computational cost and further simplifying the optimization problem to yield the final formulation of the NPV-DeePC problem as follows:
\begin{subequations} \label{eq:NpvDeePC}
\begin{align}  
    \min_{\Xi}&~ J(u,y) + \lambda_g \| \tilde{\mathbf{g}} (k) \|^2,\\
    \text{s.t.}& 
    \begin{bmatrix}
        \boldsymbol{\Upphi}_{HL}\\
        \mathbf{1}^\mathsf{T}
    \end{bmatrix}
    \mathcal{Y}_f^\dagger\tilde{\mathbf{g}}(k)=\boldsymbol{0}, \label{eq:NPV-DeePC_kernel}\\
    &\mathcal{Y}_f
    \begin{bmatrix}
        \boldsymbol{\Upphi}_{HL}\\
        \mathbf{1}^\mathsf{T}
    \end{bmatrix} ^\dagger
    \begin{bmatrix}
         \upphi_{HL}(k)\\
          1
    \end{bmatrix}
    +\mathbf{\tilde{g}}(k) = \mathbf{y}(k),\label{eq:NPV-DeePC_NLSconst}\\
    &(\mathbf{u}(k), \mathbf{y}(k))\in \mathbb{U}^N\times \mathbb{Y}^N\label{eq:NPV-DeePC_admconst}.
\end{align}
\end{subequations}
where $\Xi := \text{col}(\mathbf{u}(k), \mathbf{y}(k), \tilde{\mathbf{g}}(k))$ represents the collection of all optimization variables. \textcolor{black}{While the dynamic constraints \eqref{eq:NPV-DeePC_kernel} and \eqref{eq:NPV-DeePC_NLSconst} are always feasible due to the trivial solution $\mathbf{\tilde{g}}(k)=\textbf{0}$, this least-squares prediction can be suboptimal in the presence of noise. A strict equality can prevent the optimizer from finding a more descriptive, non-zero $\mathbf{\tilde{g}}(k)$. Therefore, to improve the numerical robustness and allow the model to find a higher-quality solution, we incorporate a slack variable into the kernel constraint \eqref{eq:NPV-DeePC_kernel} and penalize it in the cost function. This provides the necessary flexibility to absorb inconsistencies caused by noise.}

The computational complexity of solving \eqref{eq:NpvDeePC} can be evaluated by the number of decision variables and constraints involved.  Specifically, the problem includes $(n_u+2n_y)N+n_l+1$ decision variables, $n_yN+n_l+1$ equality constraints, and $2(n_u+n_y)N$ inequality constraints.  Notably, the complexity scales with the prediction horizon $N$ and the  number of neurons in the final hidden layer of the target network $n_l$.

\section{Simulation Results and Validation}
To evaluate the performance of the proposed control strategy for the APPJ system, we conducted multiple simulations under two distinct case studies: reference surface temperature tracking and optimal thermal dose delivery. These scenarios were designed to capture the system's behavior under varying operating conditions while enforcing input and output constraints outlined in TABLE \ref{tb:const} to ensure reliable and practical operation. 
A validated physics-based model of \textcolor{black}{a radio-frequency (RF) APPJ in pure} Argon, as detailed in \cite{gidon2017effective}, served as the plant model. \textcolor{black}{The simulations were executed with a sampling time of $\delta t = 0.5$ sec, chosen to balance computational efficiency with the need to capture the transient dynamics of the plasma jet. These simulations were run} on a PC with an Intel Core i7-13700 3.40 GHz CPU and 32 GB of RAM, utilizing the Python interface to the open-source \texttt{CasADi} library \cite{Andersson2018} with \texttt{IPOPT} as the solver.

\begin{table}
\caption{Input and state constraints of the APPJ.}
\begin{center}
\begin{tabular}{|c|c|c|c|}
\hline
\textbf{Variable} & \textbf{Unit} & \textbf{Lower Limit} & \textbf{Upper Limit} \\
\hline
$P$ & W & 1.5 & 8.0 \\
\hline
$q$ & slm & 1.0 & 6.0 \\ 
\hline
$T_s$ & $^{\circ}$C & 25 & 42.5 \\ 
\hline
$T_g$ & $^{\circ}$C & 20 & 80 \\
\hline
\end{tabular}
\label{tb:const}
\end{center}
\end{table}

\subsection{Data Generation and Preprocessing}
\begin{figure}
\centering
\subfloat[Open-loop inputs.]{\includegraphics[width=\columnwidth]{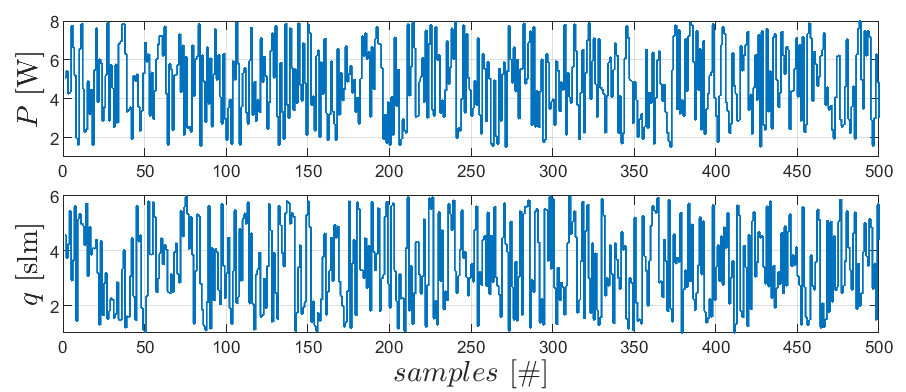}\label{fig:OpenLoop_input}}
\vfil
\subfloat[Tip-to-surface distance data.]{\includegraphics[width=\columnwidth]{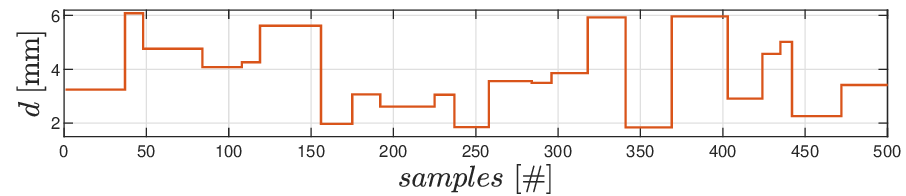}\label{fig:OpenLoop_distance}}
\vfil
\subfloat[Open-loop outputs.]{\includegraphics[width=\columnwidth]{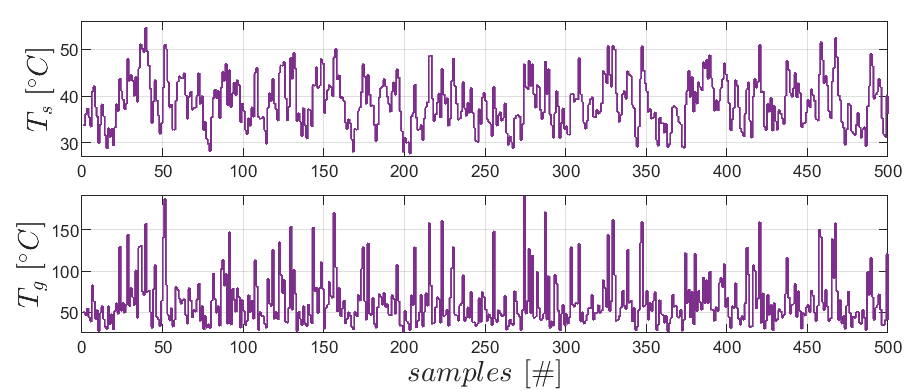}\label{fig:OpenLoop_output}}
\caption{Dataset used for HyperDNN training, showing a representative 500-sample subset.}
\label{fig:OpenLoop_data}
\end{figure}

To generate a rich dataset for offline HyperDNN training and \textcolor{black}{validation}, we performed open-loop simulations on the analytical model by exciting the system with uniformly random distributed inputs over \textcolor{black}{$25,000$} data points, as shown in Fig. \ref{fig:OpenLoop_input}. Meanwhile, the tip-to-surface distance was varied following a piecewise constant pattern with uniformly distributed random variations within the acceptable range, as depicted in Fig. \ref{fig:OpenLoop_distance}. The corresponding system outputs were recorded and presented in Fig. \ref{fig:OpenLoop_output}. The measurements used for constructing the dataset were assumed to be noise-free. For better visualization, a subset of 500 samples of the signals was shown in Fig. \ref{fig:OpenLoop_data}. This dataset ensures that the inputs to the neural network remain persistently exciting, allowing the weights to be accurately trained to capture the nonlinear behavior of the APPJ under varying operating conditions.

\textcolor{black}{The dataset was chronologically partitioned into training ($64\%$), validation ($16\%$), and test ($20\%$) sets to avoid data leakage and ensure unbiased evaluation of generalization. To facilitate stable training, a MinMaxScaler was applied to scale inputs $u$, tip-to-surface distance $d$, and outputs $y$ to the range $[-3,3]$, fitted solely on the training set and subsequently used to transform the validation and test sets. The scaled time-series data was arranged into Hankel matrices $\mathscr{H}$, $\mathcal{P}$, and $\mathcal{Y}_f$ with past and future horizons of $T_{\text{ini}}=5$ and $N=10$, respectively. Each column of these matrices provides a training sample: $\mathscr{H}$ provides the target network input $\mathbf{u}_{NN}$, $\mathcal{P}$ provides the hypernetwork input $\mathbf{p}$, and $\mathcal{Y}_f$ contains the corresponding output sequence $\mathbf{y}$.}

\color{black}
\subsection{HyperDNN Training and Neural Space Construction}
The neural space is constructed offline by training the HyperDNN on the generated dataset, providing the foundation for accurate output trajectory prediction. Training aims not only to minimize prediction error but also to preserve the richness of the lifted Hankels in neural space.

The training procedure employs the Adam optimizer with a learning rate of $10^{-4}$ and gradient clipping at an $\ell_2$-norm of $1.0$ to prevent exploding gradients. Training progress is monitored on the validation set using two adaptive callbacks. Early stopping terminates training when the validation loss ceases to improve by at least $10^{-6}$ within a patience window, restoring the best-performing weights. In parallel, a learning rate scheduler (ReduceLROnPlateau) reduces the learning rate by a factor of 0.5 whenever validation loss stagnates, enabling finer updates and avoiding suboptimal local minima.

To determine suitable architectures for both the target network and the hypernetwork, a systematic study was conducted across multiple configurations. The objective was to achieve sufficient representational capacity to capture the nonlinear dynamics, while avoiding over-parameterization that could cause overfitting, increase computational burden, and impose constraints on real-time control implementation. Each configuration was trained under identical settings (optimizer, learning rate, regularization), with $\tanh$ activation in all hidden layers, and the hypernetwork tasked with adapting the parameters of the final hidden layer of the target network. Performance was evaluated on the test data using the Mean Squared Error ($MSE$) and the Best Fit Rate ($BFR$) defined as
\begin{equation}
    BFR = \max \left( 1- \frac{\|\mathbf{y}_i-\hat{\mathbf{y}}_i\|}{\|\mathbf{y}_i-\bar{\mathbf{y}}\|} ,0\right)\times 100\%,
\end{equation}
where $\hat{\mathbf{y}}_i$ is the $i$-th predicted output trajectory and $\bar{\mathbf{y}}$ is the mean of the true sequence. Additional criteria such as the condition number $\kappa$ and minimum singular value $\varsigma_{\min}$ of $\text{col}(\boldsymbol{\Upphi}_{HL}, \boldsymbol{1}^\mathsf{T})$ were also included to evaluate the numerical quality of the constructed neural space. The performance of different architectures is summarized in Table~\ref{tb:Performance}, from which Configuration~3 (\textbf{C3}) was selected as the final design due to its favorable trade-off between accuracy, conditioning, and computational efficiency.  

\begin{table}
\textcolor{black}{
\centering
\caption{Performance results for various network configurations.}
\label{tb:Performance}
\footnotesize
\begin{tabular}{|l|c|c|c|c|c|}
\hline
\textbf{Parameter} & \textbf{C1} & \textbf{C2} & \textbf{C3} & \textbf{C4} & \textbf{C5} \\
\hline
\multicolumn{6}{|c|}{\textit{Main Network}} \\
\hline
Layers & 1 & 2 & 2 & 3 & 3 \\
Neurons & [32] & [64,32] & [64,32] & [64,32,32] & [128,64,32] \\
\hline
\multicolumn{6}{|c|}{\textit{Hypernetwork}} \\
\hline
Layers & 1 & 1 & 1 & 2 & 2 \\
Neurons & [16] & [16] & [32] & [32,16] & [32, 16] \\
\hline
\multicolumn{6}{|c|}{\textit{Performance}} \\
\hline
$MSE$ & 3.61 & 1.63 & 1.35 & 1.76 & 2.14 \\
$BFR$ (\%) & 89.64 & 92.13 & 92.79 & 91.85 & 91.16 \\
$\kappa$ & 23.05 & 2.14 & 2.38 & 2.11 & 2.77 \\
$\varsigma_{\min}$ & 1.83 & 1.29 & 1.26 & 0.99 & 1.12 \\
\hline
\end{tabular}}
\end{table}

\color{black}
\subsection{Control Performance}
The effectiveness of the proposed NPV-DeePC for APPJs was evaluated through simulations on reference surface temperature tracking and thermal dose delivery. Its performance was benchmarked against three controllers: neural DeePC \cite{lazar2024neural}, standard DeePC \cite{coulson2019data}, and MPC, each configured with the same past and prediction horizon. Due to DeePC’s high computational complexity compared to the other methods, $L = 300$ data points were used to construct its Hankel matrices. For MPC design, \textcolor{black}{similar to \cite{gidon2017model},} an LTI model obtained via subspace identification was employed.

\begin{figure}
\centering
\includegraphics[width=\columnwidth]{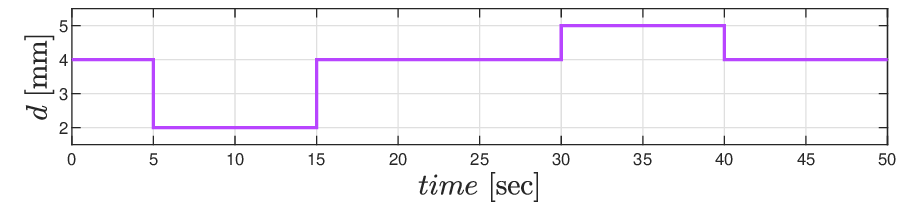}
\caption{\textcolor{black}{Tip-to-surface distance variation for the reference tracking experiment.}}
\vspace{-15pt}
\label{fig:d_trk}
\end{figure}

The first experiment evaluates the controllers' capability to track the surface temperature reference under varying conditions. To evaluate robustness, simulations were conducted both with and without measurement noise. In the noisy scenario, zero-mean Gaussian noise with a standard deviation of $0.2~^{\circ}\text{C}$ was added. Furthermore, the controllers' adaptability to varying operating conditions was tested by dynamically altering the APPJ's tip-to-surface distance, as shown in Fig. \ref{fig:d_trk}.

\begin{figure*}
\includegraphics[width=\textwidth]{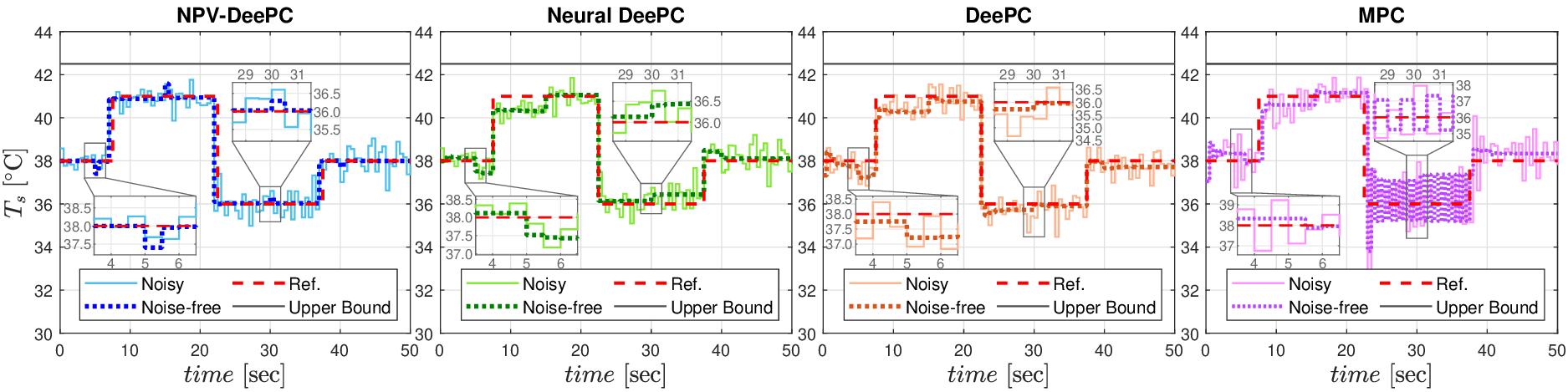}
\caption{Tracking performance of the proposed controller compared to the benchmarks for noise-free and noisy measurements.}
\label{fig:outputTracking}
\vspace{10pt}
\includegraphics[width=\textwidth]{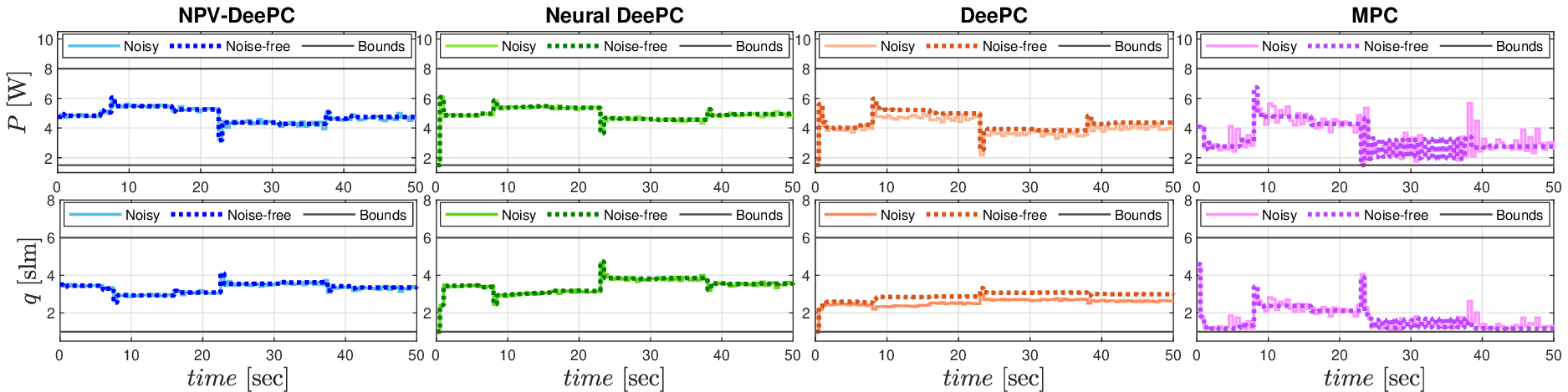}
\caption{Comparison of control inputs for reference tracking experiment under noise-free and noisy measurements.}
\label{fig:inputTracking}
\end{figure*}

The tracking performance of all controllers is presented in Fig. \ref{fig:outputTracking}. While each maintained the surface temperature within an acceptable range, their adaptability varied significantly. Evidently, the NPV-DeePC method was the only controller capable of effectively adjusting to the dynamically changing tip-to-surface distance, successfully tracking the desired trajectory. In contrast, the other controllers exhibited noticeable 
\textcolor{black}{tracking errors under varying conditions.} 

The corresponding control effort, shown in Fig. \ref{fig:inputTracking}, demonstrates that all controllers successfully kept the control inputs within the allowable range. Notably, the data-driven controllers exhibited smooth and consistent control actions, whereas the MPC showed pronounced oscillations in both electric power and gas flow rate, struggling to track the desired trajectory.

\textcolor{black}{To quantitatively evaluate the tracking performance of the controllers, three performance indices were computed over $n_{\text{sim}}$ simulation steps: Root Mean Square Error ($RMSE$), Integral Square Error ($ISE$), and control energy ($J_u$). $RMSE$ measures the average magnitude of the tracking error, $ISE$ quantifies the cumulative squared tracking error, and $J_u$ evaluates the total control effort. These indices are defined as:}
\begin{align}\label{eq:index}
        \nonumber &RMSE = \sqrt{\frac{\sum_{k=T_{\text{ini}}}^{n_{\text{sim}}} \| y(k) - r(k) \|^2}{n_{\text{sim}} - T_{\text{ini}} + 1}},\\ &ISE=\sum_{k=T_{\text{ini}}}^{n_{\text{sim}}}\|y(k)-r(k)\|^2,\quad        
        J_u =\sum_{k=T_{\text{ini}}}^{n_{\text{sim}}}\|u(k)\|^2.
\end{align}

Furthermore, the computational complexity of the controllers was assessed via the mean CPU time ($\bar{t}_{CPU}$) required to compute control actions. This metric offers insights into their practicality and real-time applicability.

\begin{figure*}
\centering
\includegraphics[width=2\columnwidth]{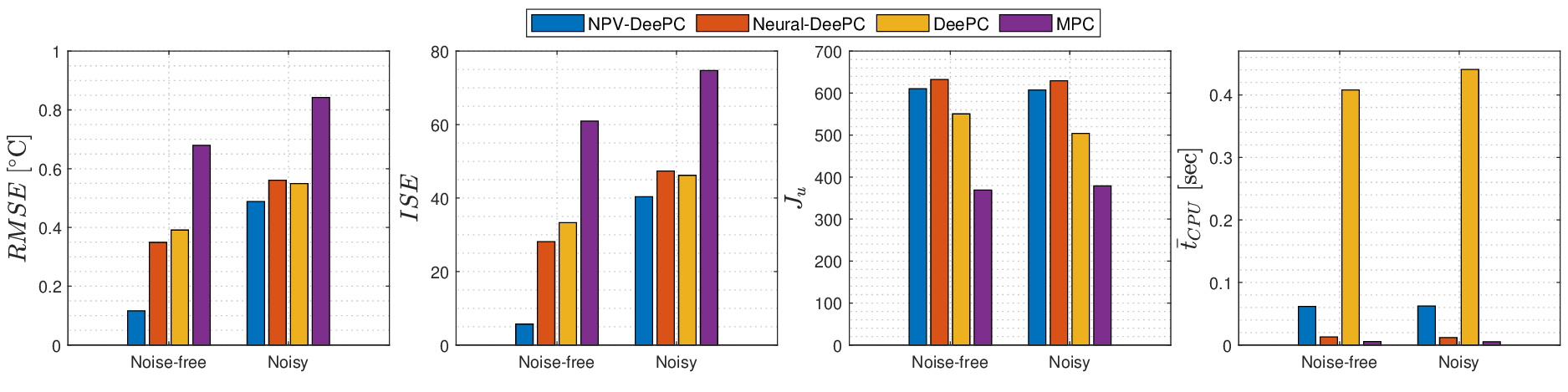}
\caption{Performance indices of the reference tracking experiment.}
\label{fig:indx_trk}
\end{figure*}

The performance indices of the controllers are presented in Fig. \ref{fig:indx_trk}. As anticipated, the proposed NPV-DeePC outperformed the other controllers, achieving an RMSE of \textcolor{black}{$0.12~^{\circ}\text{C}$} in the noise-free scenario and \textcolor{black}{$0.49~^{\circ}\text{C}$} under measurement noise. In contrast, the MPC controller had the poorest tracking performance, with an RMSE of $0.68~^{\circ}\text{C}$ in the noise-free case and $0.84~^{\circ}\text{C}$ under noise. A similar trend was observed in the ISE results, where NPV-DeePC maintained lower values of \textcolor{black}{$5.69$ and $40.34$} in \textcolor{black}{noise-free and noisy} conditions, respectively, while MPC exhibited significantly higher values of 
$60.97$ and $74.70$ in the corresponding cases.

The comparison of the control energy index shows that NPV-DeePC achieves superior tracking while consuming energy comparable to other data-driven methods. In contrast, the MPC controller, despite consuming $40\%$ less control energy, failed to match the tracking performance of the other controllers.
\textcolor{black}{Regarding computational cost, NPV-DeePC achieved a mean CPU time of approximately $\bar{t}_{CPU}=60~\text{ms}$, which, while higher than Neural DeePC ($13~\text{ms}$), remained significantly lower than standard DeePC. Importantly, this moderate overhead enabled NPV-DeePC to deliver substantially improved tracking accuracy, while maintaining practical real-time feasibility. By contrast, MPC achieved the lowest computation time of $5~\text{ms}$ but at the expense of degraded tracking performance.}

\begin{figure}
\centering
\includegraphics[width=\columnwidth]{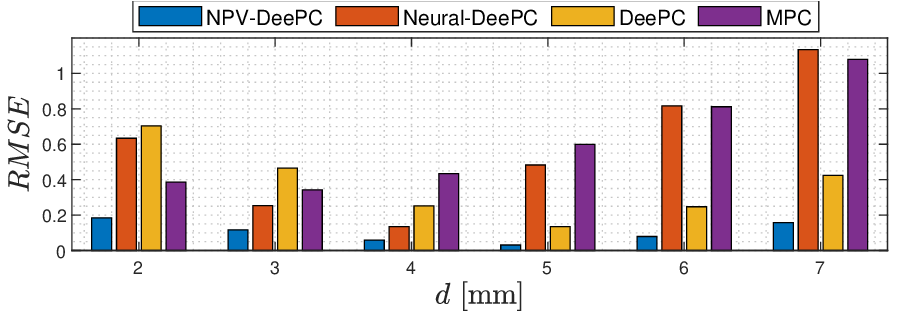}
\caption{Effect of tip-to-surface distance on RMSE index.}
\label{fig:RMSE}
\end{figure}

A closer analysis of the DeePC results reveals that when the tip-to-surface distance increases to 5 mm between $t=30$ sec and $t=40$ sec, tracking performance improves, yielding negligible error. To further investigate the impact of tip-to-surface distance on control performance, a series of simulations were conducted under noise-free conditions, each with a fixed distance ranging from $2$ to $7$ mm. The resulting RMSE values, shown in Fig. \ref{fig:RMSE}, demonstrate that NPV-DeePC consistently achieved the best performance across all distances. Furthermore, as the distance increased beyond 4 mm, DeePC generally outperformed neural DeePC and achieved its best performance at $d = 5$ mm, suggesting increasingly linear system behavior at larger distances. However, the performance of neural DeePC can be further enhanced by leveraging a physics-guided subspace neural network architecture, as proposed in \cite{lazar2023nonlinear}. This architecture integrates a parallel linear model with a neural network, enabling the model to effectively learn the underlying linear behavior.

\begin{figure}
\centering
\includegraphics[width=\columnwidth]{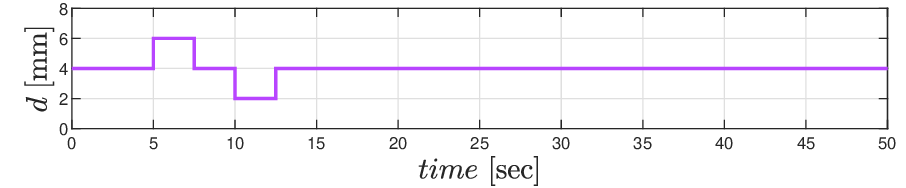}
\caption{Tip-to-surface distance for the thermal dose delivery experiment.}
\label{fig:d_cem}
\end{figure}

The second experiment aimed to evaluate the optimal thermal dose delivery performance of the APPJ to the target surface. In this scenario, the control objective is to deliver the target thermal dose, denoted as $\text{CEM}_{T}$, to the substrate. Therefore, the controller cost function includes only the terminal cost, $J = \| \text{CEM}_{T} - \text{CEM}(N) \|_2^2$, where $\text{CEM}(k)$ is the current delivered dose and $\text{CEM}(N)$ represents the predicted CEM at the end of the prediction horizon ($N=5$). The impact of variations in the tip-to-surface distance on thermal dose delivery was assessed by introducing a perturbation, as shown in Fig. \ref{fig:d_cem}. These variations were applied prior to  
$15~\text{sec}$, during which the majority of the thermal dose is delivered. This timing ensured that the controllers had sufficient opportunity to respond and effectively adjust to the new operating condition.

\begin{figure*}
\includegraphics[width=\textwidth]{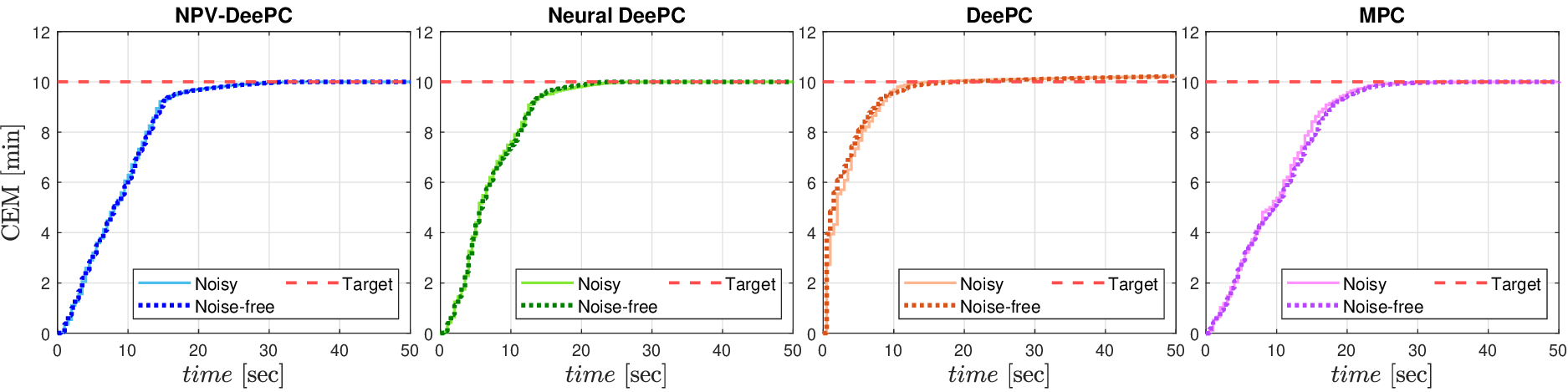}
\caption{Thermal dose delivery of the proposed controller compared to the benchmarks for noise-free and noisy measurements.}
\label{fig:CEMdelivery}
\vspace{10pt}
\includegraphics[width=\textwidth]{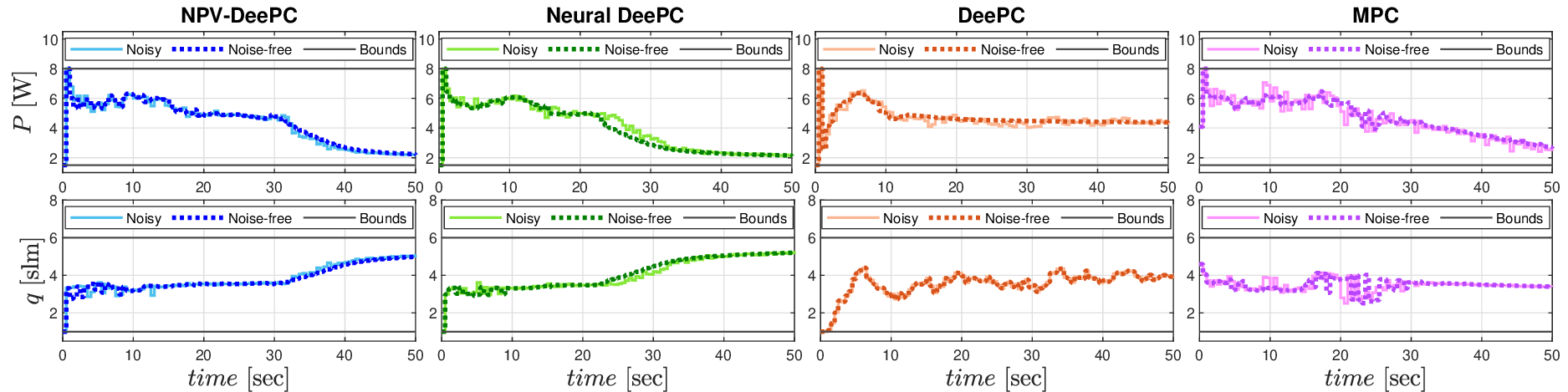}
\caption{Control inputs for the thermal dose delivery scenario of the proposed controller compared to the benchmarks for noise-free and noisy measurements.}
\label{fig:inputCem}
\end{figure*}

The CEM thermal dose delivery results under noise-free and noisy measurement conditions are presented in Fig. \ref{fig:CEMdelivery}. It can be observed that nearly all controllers effectively delivered the specified thermal dose to the substrate, with the exception of the DeePC controller, which exceeded the target, \textcolor{black}{raising a safety concern due to thermal overexposure, that may cause permanent and irreversible changes to the substrate’s physical properties.} Notably, due to the adaptive nature of the NPV-DeePC, it was able to maintain a consistent rate of thermal dose delivery, while the other controllers experienced variations in the delivery rate. This was achieved with reasonable control effort, as depicted in Fig. \ref{fig:inputCem}. These results highlight the effective performance of NPV-DeePC in ensuring stable, controlled thermal dose delivery, thereby guaranteeing both substrate safety and treatment efficacy.

\section{Conclusion}
This paper introduced the Neural Parameter-Varying Data-enabled Predictive Control (NPV-DeePC) framework to address the challenges of controlling Atmospheric Pressure Plasma Jets (APPJs) in biomedical applications. By integrating a Hyper Deep Neural Network (HyperDNN) with the neural DeePC, the proposed method enables adaptive representation of time-varying dynamics, ensuring reliable and high-precision performance across diverse operating conditions.

Comprehensive simulations benchmarked NPV-DeePC against neural DeePC, standard DeePC, and MPC across two key tasks: surface temperature tracking and optimal thermal dose delivery. The evaluation considered both noise-free and noisy environments, as well as variations in the jet’s tip-to-surface distance. The results showed that NPV-DeePC consistently achieved superior tracking accuracy and adaptability. In particular, it maintained stable performance under dynamic variations, effectively enforced input and state constraints, and delivered more accurate and consistent thermal doses compared to alternative controllers. Importantly, these benefits were obtained with moderate computational cost, supporting the feasibility of real-time implementation.

Overall, the findings highlight NPV-DeePC as an effective solution for APPJs in sensitive biomedical treatments, where precision and reliability are paramount. More broadly, the approach provides a generalizable paradigm for controlling other nonlinear and parameter-varying systems where conventional approaches face scalability or adaptability limitations. Future work will aim to establish formal stability and recursive feasibility guarantees, and to further enhance robustness \textcolor{black}{against noise and disturbances.}


\bibliographystyle{IEEEtran}
\bibliography{2-Ref.bib} 

\end{document}